\documentclass{cernyrep}

\begin{document}
\title{Study of the $pp \rightarrow np\pi^{+}$ reaction at 1.25 GeV with HADES}

\author
{
T.~Liu\footnote{\begin{normalsize}liu@ipno.in2p3.fr\end{normalsize}}$^{\,13}$,
G. Agakishiev$^{\,8}$, 
C.~Agodi$^{\,1}$,
A.~Balanda$^{\,3,e}$,
G.~Bellia$^{\,1,a}$,
D.~Belver$^{\,15}$,
A.~Belyaev$^{\,6}$,
A.~Blanco$^{\,2}$,
M.~B\"{o}hmer$^{\,11}$,
J.~L.~Boyard$^{\,13}$,
P.~Braun-Munzinger$^{\,4}$
% ...
}

\vspace{5cm}

\institute
{
~\\
~\\
$^1$Istituto Nazionale di Fisica Nucleare - Laboratori Nazionali del Sud, 95125~Catania, Italy,\\
$^{2}$LIP-Laborat\'{o}rio de Instrumenta\c{c}\~{a}o e F\'{\i}sica Experimental de Part\'{\i}culas , 3004-516~Coimbra, Portugal\\
$^3$Smoluchowski Institute of Physics, Jagiellonian University of Cracow, 30-059~Krak\'{o}w, Poland,\\
$^4$GSI Helmholtzzentrum f\"{u}r Schwerionenforschung, 64291~Darmstadt, Germany,\\
$^5$Institut f\"{u}r Strahlenphysik, Forschungszentrum Dresden-Rossendorf, 01314~Dresden, Germany,\\
$^6$Joint Institute of Nuclear Research, 141980~Dubna, Russia,\\
$^7$Institut f\"{u}r Kernphysik, Johann Wolfgang Goethe-Universit\"{a}t, 60438 ~Frankfurt, Germany,\\
$^8$II.Physikalisches Institut, Justus Liebig Universit\"{a}t Giessen, 35392~Giessen, Germany,\\
$^9$Istituto Nazionale di Fisica Nucleare, Sezione di Milano, 20133~Milano, Italy,\\
$^10$Institute for Nuclear Research, Russian Academy of Science, 117312~Moscow, Russia,\\
$^{11}$Physik Department E12, Technische Universit\"{a}t M\"{u}nchen, 85748~M\"{u}nchen, Germany,\\
$^{12}$Department of Physics, University of Cyprus, 1678~Nicosia, Cyprus,\\
$^{13}$Institut de Physique Nucl\'{e}aire (UMR 8608), CNRS/IN2P3 - Universit\'{e} Paris Sud, F-91406~Orsay Cedex, France,\\
$^{14}$Nuclear Physics Institute, Academy of Sciences of Czech Republic, 25068~Rez, Czech Republic,\\
$^{15}$Departamento de F\'{\i}sica de Part\'{\i}culas, Univ. de Santiago de Compostela, 15706~Santiago de Compostela, Spain,\\
$^{16}$Instituto de F\'{\i}sica Corpuscular, Universidad de Valencia-CSIC, 46971~Valencia, Spain,\\
$^a$Also at Dipartimento di Fisica e Astronomia, Universit\`{a} di Catania, 95125~Catania, Italy,\\
$^b$Also at ISEC Coimbra, ~Coimbra, Portugal,\\
$^c$Also at Technische Universit\"{a}t Dresden, 01062~Dresden, Germany,\\
$^d$Also at Dipartimento di Fisica, Universit\`{a} di Milano, 20133~Milano, Italy,\\
$^e$Also at Panstwowa Wyzsza Szkola Zawodowa , 33-300~Nowy Sacz, Poland.
}
\maketitle

\newpage

\begin{abstract}
In $pp$ collisions at 1.25 GeV kinetic energy, the HADES collaboration aimed at investigating the di-electron production related to $\Delta$ (1232) Dalitz decay ($\Delta^{+} \rightarrow pe^{+}e^{-}$). In order to constrain the models predicting the cross section and the production mechanisms of $\Delta$ resonance, the hadronic channels have been measured and studied in parallel to the leptonic channels. The analyses of $pp\rightarrow np\pi^{+}$ and $pp\rightarrow pp\pi^{0}$ channels and the comparison to simulations are presented in this contribution, in particular the angular distributions being sensitive to $\Delta$ production and decay. The accurate acceptance corrections have been performed as well, which could be tested in all the phase space region thanks to the high statistic data. These analyses result in an overall agreement with the one-$\pi$ exchange model and previous data.
\end{abstract}

\section{Introduction}
The High-Acceptance Dielectron Spectrometer (HADES) is a magnetic spectrometer in operation at the heavy ion synchrotron facility SIS at GSI Darmstadt. It is designed for di-electron spectroscopy in nucleus-nucleus collisions around 1-2 AGeV to study hot and dense nuclear matter. Di-electron production in C+C collisions at 1 and 2 AGeV have been reported on \cite{CC2,CC1}. In the invariant mass region 0.15 GeV/$c^{2}$ < $M_{e^{+}e^{-}}$ < 0.5 GeV/$c^{2}$ the measured pair yield shows a strong excess above the contribution expected from hadron decays after freeze-out. In this mass range, the $\eta$ Dalitz decay is well under control, but two other di-electron sources which play an important role are poorly known: the $\Delta$ (1232) resonance Dalitz decay ($\Delta^{+} \rightarrow pe^{+}e^{-}$) and $\textit{NN}$ bremsstrahlung ($NN \rightarrow NNe^{+}e^{-}$).\\
For a better understanding of the contribution from such processes, HADES studied $pp$ and $dp$ interactions at 1.25 GeV kinetic energy \cite{pp,dp}, just below the $\eta$ production threshold. Due to the small expected contribution for the $\textit{NN}$ bremsstrahlung, the $pp$ reaction at 1.25 GeV is expected to be mostly sensitive to the $\Delta$ Dalitz decay \cite{Teis}. This process is studied by leptonic inclusive ($pp\rightarrow e^{+}e^{-}X$) and exclusive ($pp\rightarrow ppe^{+}e^{-}$) channel analyses. At the same time, we measured the hadronic channels ($pp \rightarrow np\pi^{+},~pp \rightarrow pp\pi^{0}$) which provide a consistency check with the analysis of leptonic channels as well as a precise and independent control of inputs (cross section, angular distribution, etc.) for $\Delta$ Dalitz decay study. The pion production is dominated by $\Delta$ excitation at $T_{\rm{kin}}=1.25$ GeV, therefore, the two hadronic channels are correlated by isospin symmetry: $\sigma (pp \rightarrow n\Delta^{++}) \sim 3\,(pp \rightarrow p\Delta^{+})$, leading to $\sigma (pp \rightarrow np\pi^{+}) \sim 5 \, (pp \rightarrow pp\pi^{0})$.  

\section{Experiment}
\begin{figure}[htb]
\centering
\includegraphics[width=6.5cm]{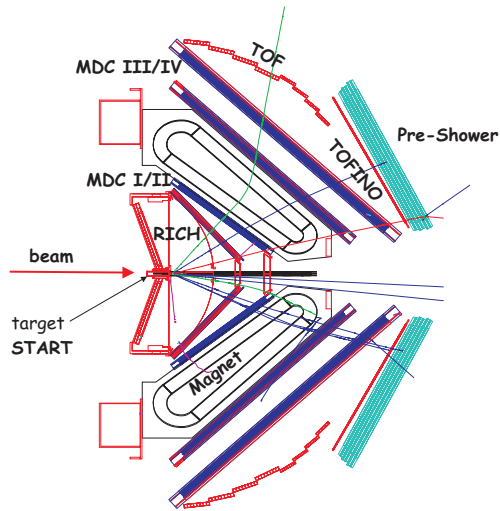}
\caption{Schematic layout of the HADES detector.}
\label{HADES_sideview}
\end{figure}
The HADES detector, as shown in Fig.~\ref{HADES_sideview}, consists of 6 identical sectors covering the full azimuthal range and polar angles from $18^{o}$ to $84^{o}$ with respect to the beam direction. Each sector contains:  A Ring Imaging CHerenkov (RICH) detector used for electron identification; two sets of Mini-Drift Chambers (MDC) with 4 modules per sector placed in front and behind the magnetic field to determine momenta of charged particles; the Time Of Flight detectors (TOF/TOFINO) and the Pre-Shower detector improving the electron identification. For reaction time measurement, a START detector is located in front of the target. An $(e^{+},e^{-})$ invariant mass resolution at the $\omega$ peak of $\sim 2.7\%$ and a momentum resolution for protons of $\sim 3\%$ can be achieved. The first level trigger is obtained by a fast multiplicity signal coming from the TOF/TOFINO wall, combined with a reaction signal from the START detector, while the second level trigger is made by using the informations from the RICH and Pre-Shower to enrich the lepton candidates signals.\\
In the April 2006 experiment, we used protons at 1.25 GeV kinetic energy, with an intensity of about $10^{7}$ particles/second, and a target filled with liquid hydrogen. The START detector was not used in this run because of the too high secondaries produced by the START detector itself. Thus a specific algorithm \cite{Marcin} has been developed to calculate the time of flight and then to identify the charged particles.

\section{Simulation}
The two hadronic channels, $pp \rightarrow np\pi^{+}$ and $pp \rightarrow pp\pi^{0}$, were simulated at a beam energy of 1.25 GeV using the PLUTO event generator \cite{pluto}. As summarized in table \ref{tab:LET}, the channels involving $\Delta$ and N* resonances are simulated with the cross sections taken from \cite{Teis}.  For the $\Delta$ resonance production, the mass distribution and angular distribution are taken from Dmitriev's calculation based on one-$\pi$ exchange model which describes very well the previous experimental data \cite{Dmitriev}. For the $\Delta$ hadronic decay, we implemented the angular differential cross section $\rm{d}\sigma/\rm{d}\Omega \sim (1+1.35 ~ \rm{cos}^{2}\theta)$ which is in agreement with the anisotropies measured by previous experiments \cite{wicklund,bacon} ($\theta$ is the $\pi$ polar angle in $\Delta$ frame with respect to the direction of momentum transfer from $N$ to $\Delta$ in the total center of mass). Theoretical predictions \cite{shuber} have been used for the angular distribution of N* production since no data exist. Simulated events were filtered through HADES acceptance to compare with the data. 
\begin{table}[h]
\begin{center}
\caption{Employed hadronic channels for the $pp$ reaction at 1.25 GeV.}
\label{tab:LET}
\begin{tabular}{p{3cm}ccc}
\hline\hline
\textbf{Outgoing channel}        & \textbf{Cross section}   & \textbf{Production processes}   &\textbf{Corresponding cross section} \\
\hline
$pp \rightarrow np\pi^{+}$       &$19.4$~mb                 &$pp \rightarrow n\Delta^{++}$    &$17.0$~mb\\
                                 &                          &$pp \rightarrow p\Delta^{+}$     &$1.9$~mb\\
                                 &                          &$pp \rightarrow pN^{*}$          &$0.5$~mb\\
\hline
$pp \rightarrow pp\pi^{0}$       &$4.0 $~mb                 &$pp \rightarrow p\Delta^{+}$      &$3.8 $~mb\\
                                 &                          &$pp \rightarrow pN^{*}$          &$0.2 $~mb\\
\hline\hline
\end{tabular}
\end{center}
\end{table}

\vspace{-0.2in}

\section{Results}
\subsection{$pp \rightarrow np\pi^{+}$ channel}
\begin{figure}[t]
\centering
\includegraphics[width=2.5in]{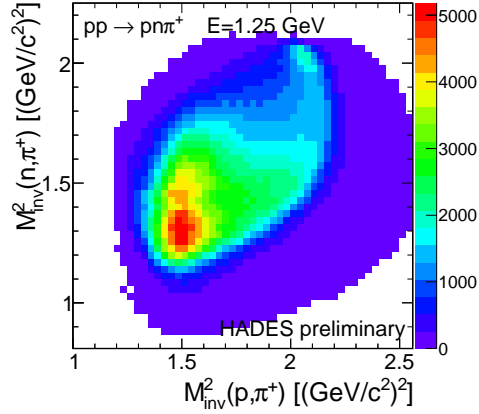}
\vspace{-0.15in}
\caption{Dalitz plot of the $pp \rightarrow np\pi^{+}$ reaction: $(p,\pi^{+})$ against $(n,\pi^{+})$ invariant mass squared distributions (preliminary HADES data).}
\label{Dalitz}
\end{figure}

The $pp \rightarrow np\pi^{+}$ channel is studied using a reconstruction of the undetected neutron. The reaction was selected first by the charged particle identification based on momentum and reconstructed time of flight \cite{tech}, then a ($p,\pi^{+}$) missing mass cut was imposed around the neutron mass. This cut efficiently suppresses the background coming from misidentified protons and two-$\pi$ contributions. 
In the Dalitz plot (Fig~.\ref{Dalitz}), one can clearly see the $\Delta^{++}$ signal located around $M_{\rm inv}^{2}(p,\pi^{+})=1.5$ (GeV/$c^{2}$)$^{2}$ corresponding to the squared mass of $\Delta^{++} (1232)$, while the $\Delta^{+}$ signal located around $M_{\rm inv}^{2}(n,\pi^{+}) = 1.5$ (GeV/$c^{2}$)$^{2}$ is less pronounced. The spot appearing in the Dalitz plot for ($n,\pi^{+}$) and ($p,\pi^{+}$) invariant masses squared around 2 (GeV/c$^{2}$)$^{2}$ is due to the $pn$ Final State Interaction (FSI). This is confirmed by a simulation where $pn$ FSI was implemented using the Jost function formalism \cite{FSI}. 

\begin{figure}[h]
\centering
 \includegraphics[width=2.5in]{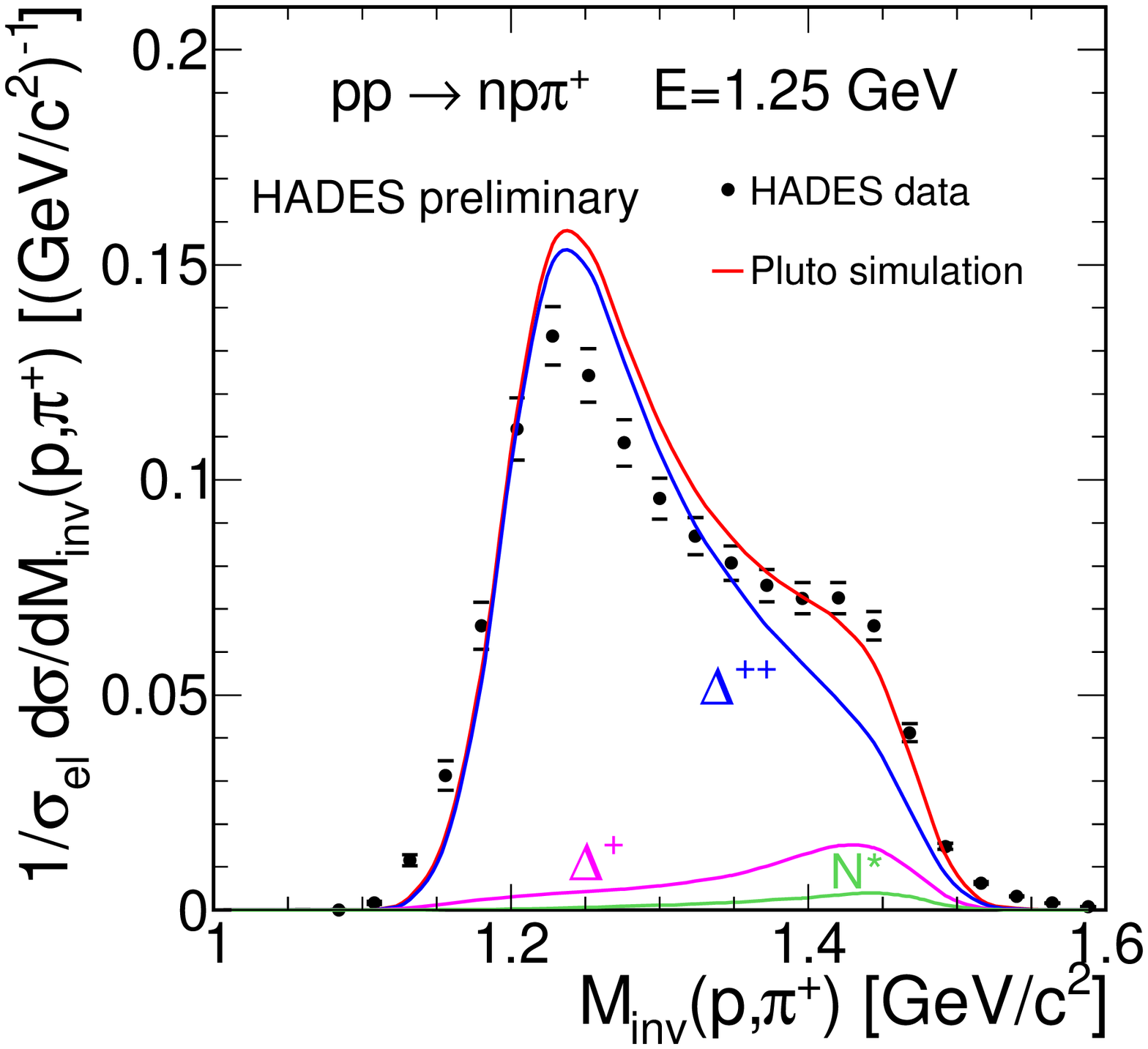}
\hspace{0.2in}
 \includegraphics[width=2.5in]{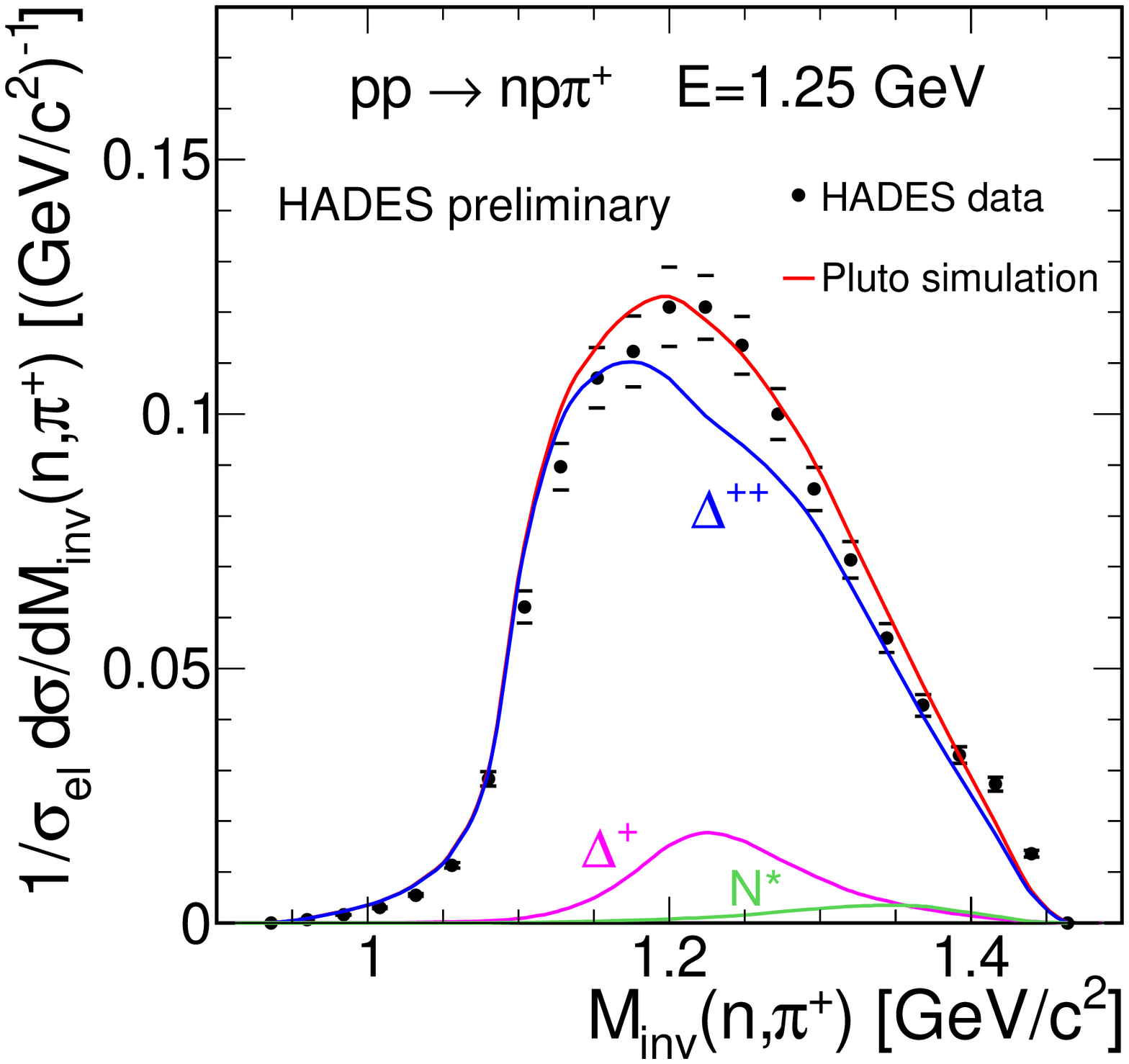}
\vspace{-0.2in}
\caption{\textbf{Left:} ($p,\pi^{+}$) and \textbf{Right:} ($n,\pi^{+}$) invariant mass distributions compared to Pluto simulation: total contribution (red), $\Delta^{++}$ (blue),  $\Delta^{+}$ (pink) and N* (green). Both the data and simulations are normalized to the total $pp$ elastic cross section.}
\label{invmass}
\end{figure}

Fig.~\ref{invmass} exhibits the projection on the ($p,\pi^{+}$) (Left) and ($n,\pi^{+}$) (Right) invariant masses and a comparison to simulation. The data and simulation are both normalized to the total $pp$ elastic cross section which is also measured in this experiment. Error bars include statistical and systematic errors due to event selection ($5\%$) and correction of trigger condition which has been applied for some of the bins. The uncertainty on the normalization to $pp$ elastic scattering is also considered as a source of systematic error ($6\%$) but not included in the error bars here. The prominent peak of $M_{\rm inv}(p,\pi^{+})$ around 1.23 GeV/$c^{2}$ confirms that most of the $\pi^{+}$ are produced via $\Delta^{++}$ decay, which is consistent with the resonance model. $N^{*}$ contribution seems also reasonable and the invariant mass distributions are rather well reproduced by the $pp \rightarrow n\Delta^{++}$ and $pp \rightarrow p\Delta^{+}$ simulations. 

\begin{figure}[h]
\centering
 \includegraphics[width=2.5in]{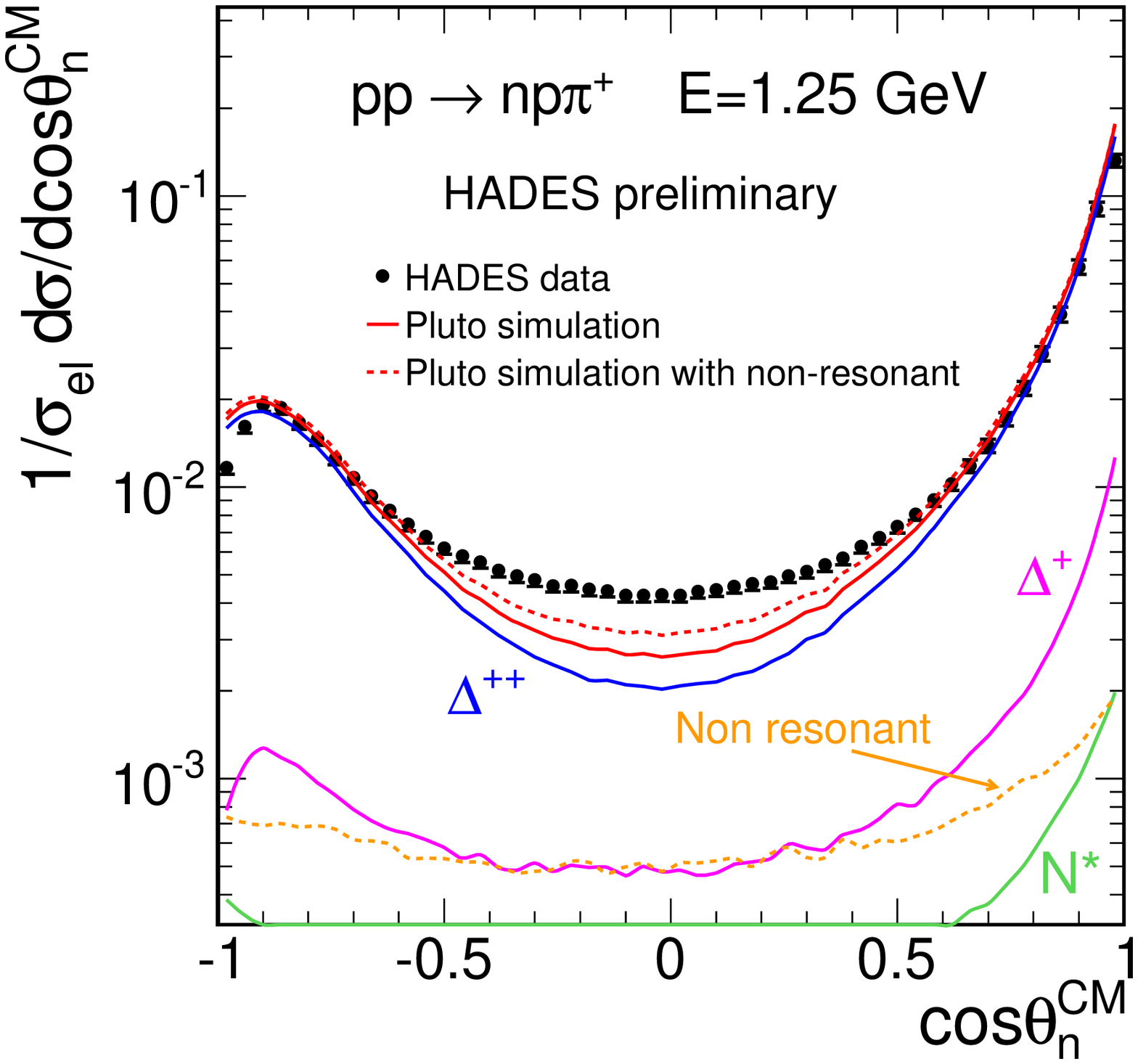}
\hspace{0.2in}
 \includegraphics[width=2.5in]{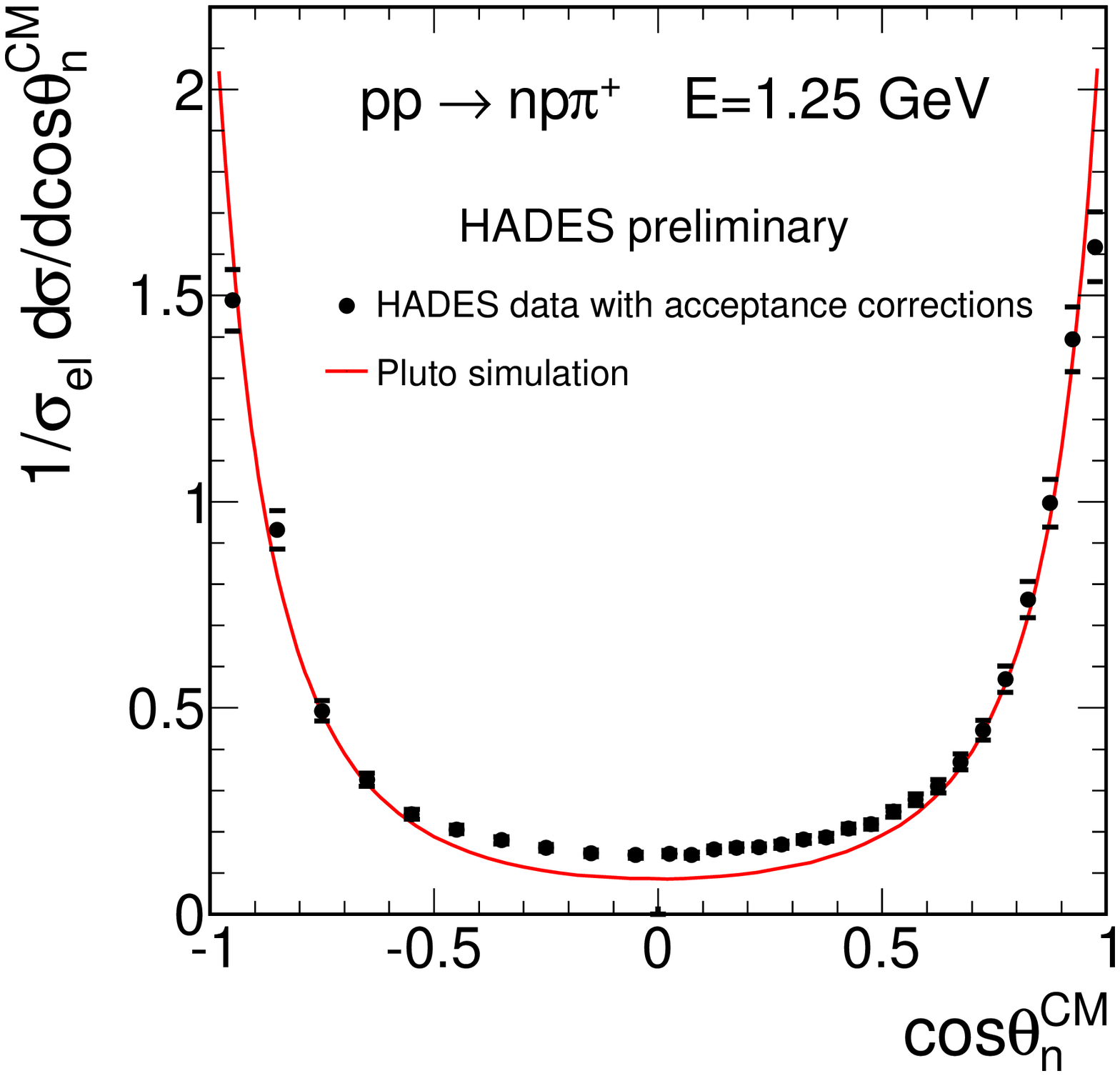}
\vspace{-0.2in}
\caption{\textbf{Left:} Angular distribution of neutron in the $pp$ center of mass system. Data (black points) compared to simulations. (Pluto simulation (red solid line) including $\Delta^{++}$ (blue), $\Delta^{+}$ (pink) and N* (green); improved Pluto simulation (red dashed line) with additional non-resonant contribution (orange dashed line)). \textbf{Right:} Angular distribution of neutron in center of mass system after acceptance correction. Data (black points) compared to Pluto simulation (red line). Both data and simulations were normalized to the total $pp$ elastic cross section.}
\label{angn}
\end{figure}

We also looked at the neutron angular distribution in the center of mass system (Fig.~\ref{angn}) which is sensitive to the angular distribution of $\Delta$ resonance production since $pp \rightarrow n\Delta^{++}$ is the dominant process. The comparison shows a good agreement between data (black points) and simulation (red solid line) for forward and backward neutron angles, however, an excess around cos$\theta_{n}=0$ is found (our data are a factor 2 above simulation). By adding a non-resonant $\pi$ production contribution with $\sigma = 0.5$ mb the simulation (red dashed line) got closer to the data points around $\theta_{n}=0$, even though, it can not fully explain the discrepancy. On the other hand, the cross section assumed here for the non-resonant contribution seems reasonable and can not be larger since it is also constrained by invariant mass distributions. But one should mention that the majority of statistics is located at forward and backward neutron angles where the yield is in agreement with the simulation and that the overall discrepancy is less than $5\%$ of the total cross section. To minimize the model dependence, a two dimensional acceptance correction has been performed using the resonance model. For each (cos$\theta_{n}$, $M_{\rm inv}(p,\pi^{+})$) bin, the correction factors are calculated as the ratio of events from simulation in full phase space and in geometrical HADES acceptance, then they are applied to correct the data. The width of the bins is adjusted to optimize the precision of the correction. The systematical error introduced by the acceptance correction procedure is estimated to $5\%$. We finally obtain the cross section $\sigma (pp~ \rightarrow~ np\pi^{+}) = 20.4 \pm 1.9$ mb, in good agreement with previous data \cite{Teis}.

\begin{figure}[h]
\centering
 \includegraphics[width=2.5in]{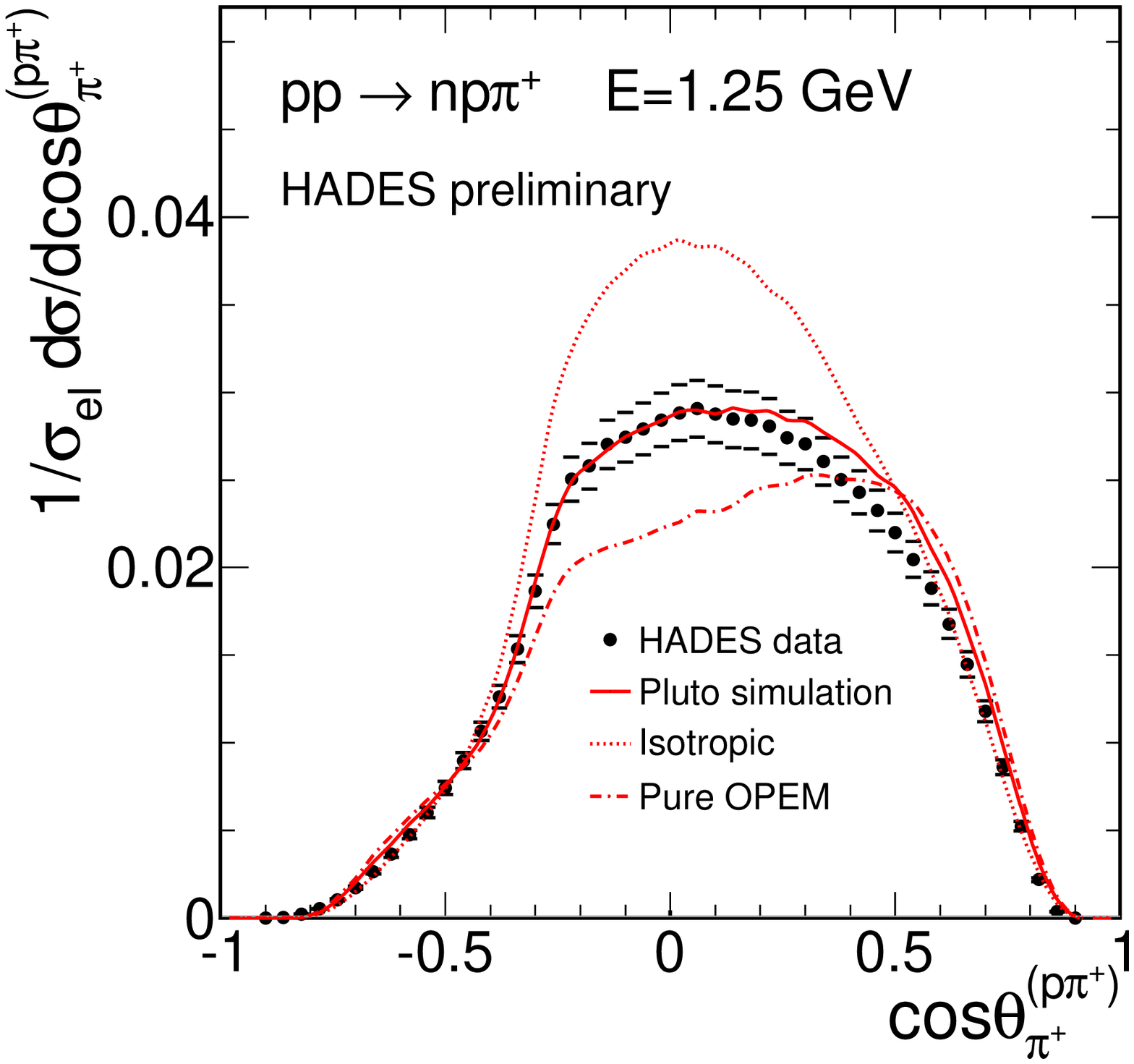}
\hspace{0.2in}
 \includegraphics[width=2.5in]{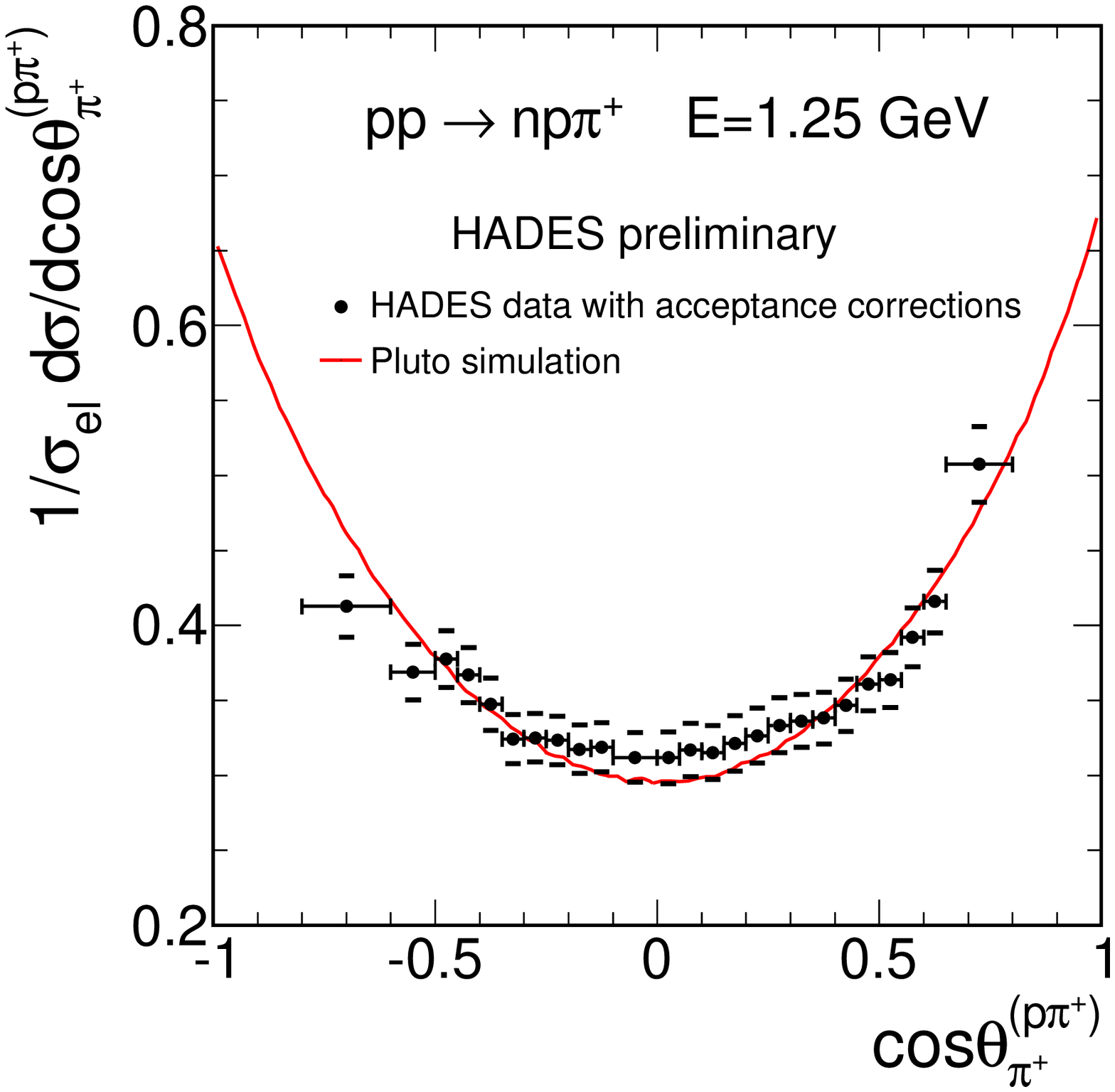}
\vspace{-0.2in}
\caption{\textbf{Left:} Angular distribution of $\pi^{+}$ in ($p,\pi^{+}$) center of mass system. Experimental data are presented as black points and compared to simulations with different $\Delta$ decay anisotropy parameters (see text). Pluto simulation: $\textit{B}=1.35$ (red solid line), pure one-$\pi$ exchange model: $\textit{B}=3$ (red dot-dashed line) and isotropic: $\textit{B}=0$ (red dashed line). \textbf{Right:} Angular distribution of $\pi^{+}$ in ($p,\pi^{+}$) reference frame after acceptance correction. Data (black points) are compared to Pluto simulation (red line).}
\label{decay}
\end{figure}

The decay angular distribution of the $\Delta$ resonance has been studied in this channel as well by looking at the angular distribution of $\pi^{+}$ in the ($p,\pi^{+}$) center of mass system. In the simulation, the angular distribution of $\pi$ in the $\Delta$ resonance frame according to $\rm{d}\sigma/\rm{d}\Omega \sim (1+\textit{B} ~ \rm{cos}^{2}\theta)$ is assumed. In order to see the sensitivity of the $\pi$ angular distribution to the anisotropy of the $\Delta$ decay, and in addition to the decay distribution mentioned above in Pluto simulation (where $\textit{B}=1.35$), the distributions expected for pure one-$\pi$ exchange model (where $\textit{B}=3$) and isotropic decay (where $\textit{B}=0$) were simulated and filtered through the HADES acceptance. The comparison to the data (left panel of Fig. \ref{decay}) shows the great sensitivity to the anisotropy of the $\Delta$ and the good agreement with anisotropies measured in previous experiments \cite{wicklund,bacon}. The acceptance correction is done using the same procedure as for the neutron angular distribution mentioned above. The right panel of Fig. \ref{decay} shows the angular distribution of $\pi^{+}$ in the ($p,\pi^{+}$) reference frame extrapolated to full solid angle, which fits nicely with the simulation.

\subsection{$pp \rightarrow pp\pi^{0}$ channel}
\begin{figure}[h]
\centering
 \includegraphics[width=2.2in]{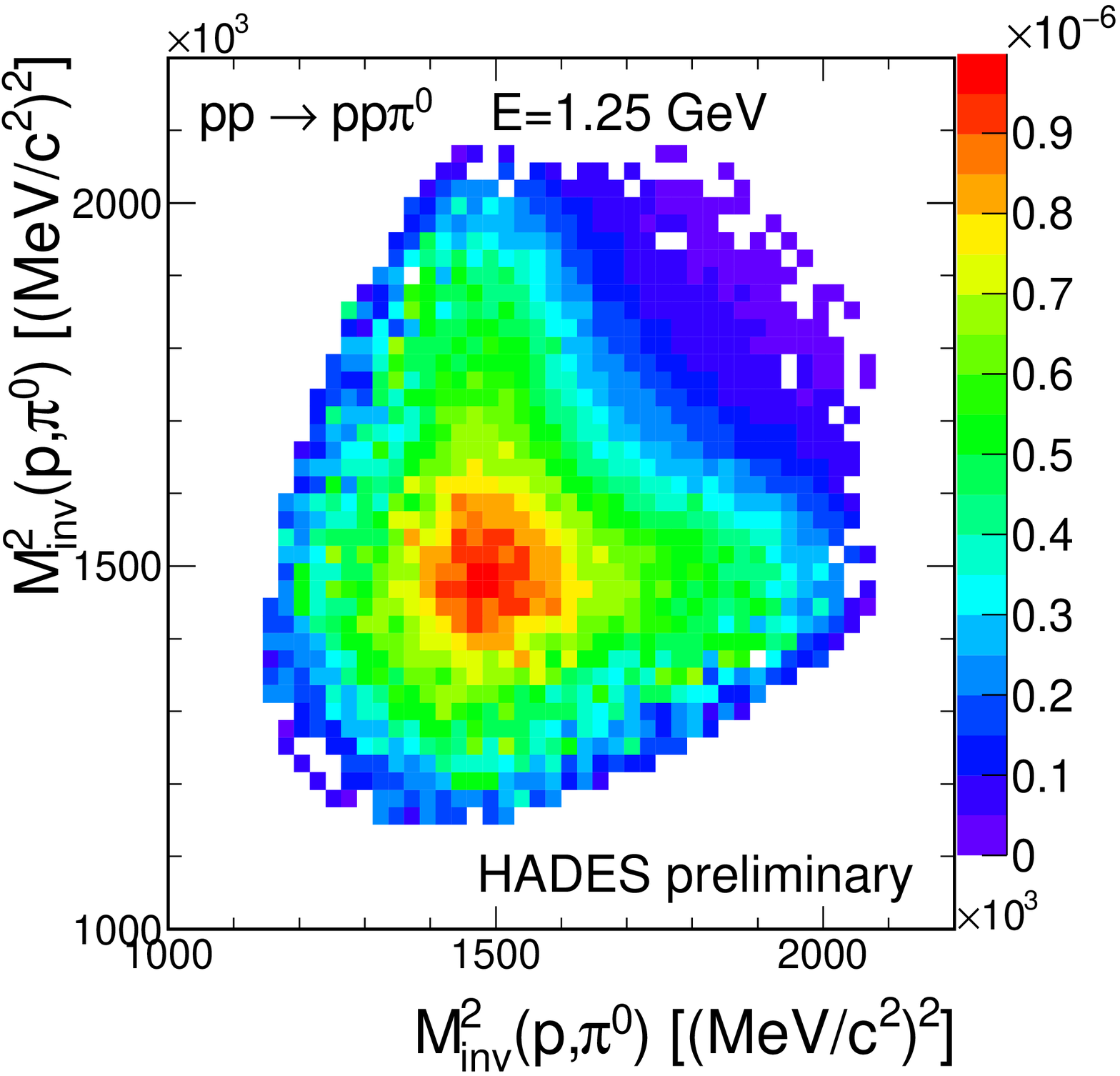}
\hspace{0.2in}
 \includegraphics[width=2.2in]{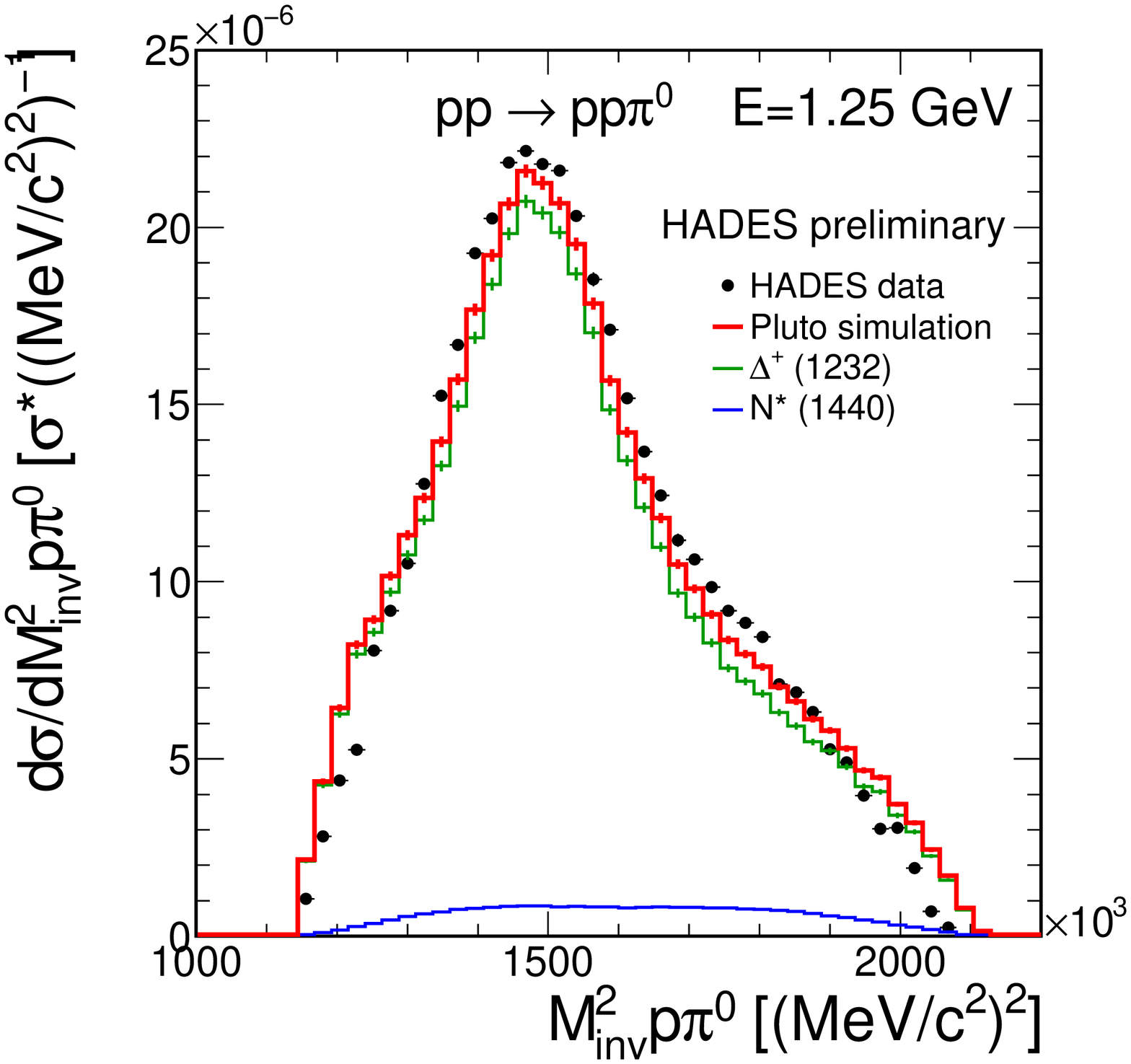}
\vspace{-0.1in}
\caption{\textbf{Left:} Dalitz plot of the $pp \rightarrow pp\pi^{0}$ reaction: $(p,\pi^{0})$ invariant mass squared distributions with preliminary HADES data.  \textbf{Right:} Projection of the Dalitz plot onto the $(p,\pi^{0})$ invariant mass squared (black points) compared to the simulation (red) assuming the $\Delta^{+}$ (green) and the N* (blue) as intermediate resonances. Both distributions are normalized to the same number of $pp$ elastic events.}
\label{pppi0}
\end{figure}

The $pp\pi^{0}$ reaction was extracted from $(p,p)$ missing mass spectrum by subtracting a background under the $\pi^{0}$ peak. The left panel of Fig. \ref{pppi0} presents the Dalitz plot of this channel. The prominent $\Delta^{+}$ signal (right panel in Fig. \ref{pppi0}) is in good agreement with the event distribution obtained from the resonance model with cross section in Table \ref{tab:LET}. This analysis results in an exclusive $\pi^{0}$ cross section extrapolated to 4$\pi$: $\sigma(pp \rightarrow pp\pi^{0}) = 4.05 \pm 0.37 $ mb, in agreement with previous data. The estimated $\sigma (pp \rightarrow np\pi^{+}) \sim 5 \, (pp \rightarrow pp\pi^{0})$ is well reproduced \cite{Teis}.

\section{Conclusion and outlook}
In summary, the preliminary HADES results for hadronic channels $pp \rightarrow np\pi^{+}$ and $pp \rightarrow pp\pi^{0}$ at 1.25 GeV have been presented. Both measured $\pi$ production cross sections agree with the previous data and the resonance model within error bars. The invariant masses and angular distributions show an overall agreement with the one-$\pi$ exchange model although a deviation is observed at large production angles. Consistent with previous data, the $\Delta$ resonance decay is clearly anisotropic. With the large statistics achieved, the spectra could be extrapolated to full phase space using 2-dimensional acceptance correction with a minimized model dependence. This provides a useful test of the resonance model used for the analysis of the di-lepton channels and is especially important for the ongoing study of $\Delta$ Dalitz decay, using $(pe^{+}e^{-})$ events.

\section*{Acknowledgements}
The HADES collaboration gratefully
acknowledges the support by BMBF grants 06MT238, 06TM970I, 06GI146I, 
06F-140, 06FY171, and 06DR135, 
by DFG EClust 153 (Germany), 
by GSI (TM-KRUE, TM-FR1, GI/ME3, 
OF/STR), 
by grants GA AS CR IAA100480803 and MSMT LC 07050 (Czech Republic), 
by grant KBN 1P03B 056 29 (Poland), by INFN (Italy), by CNRS/IN2P3 (France), 
by grants MCYT FPA2006-09154, XUGA PGID IT06PXIC296091PM and CPAN CSD2007-00042
(Spain), 
by grant FTC POCI/FP /81982 /2007  (Portugal),
by grant UCY-10.3.11.12 (Cyp\-rus), by INTAS grant
06-1000012-8861 and EU contract RII3-CT-2004-506078.

\end{document}